\documentclass{article}
\usepackage{spconf,amsmath,graphicx}
\usepackage[moderate]{savetrees}
\usepackage{pifont}
\usepackage{multirow}
\usepackage{makecell}
\usepackage{listings}
\usepackage{xcolor}
\usepackage{stix}
\usepackage{booktabs}
\usepackage{algorithm}
\usepackage{algpseudocode}

\usepackage{hyperref}
\hypersetup{
    colorlinks=true,
    linkcolor=purple,
    filecolor=magenta,      
    urlcolor=purple,
}

\newcommand{\RR}{\mathbb{R}}
\usepackage[utf8]{inputenc}

\definecolor{codegreen}{rgb}{0,0.6,0}
\definecolor{codegray}{rgb}{0.5,0.5,0.5}
\definecolor{codepurple}{rgb}{0.58,0,0.82}
\definecolor{backcolour}{rgb}{0.95,0.95,0.92}

\newcommand{\newpara}[1]{\vspace{8pt}\noindent\textbf{#1}}
\usepackage{array}
\newcommand{\PreserveBackslash}[1]{\let\temp=\\#1\let\\=\temp}
\newcolumntype{C}[1]{>{\PreserveBackslash\centering}p{#1}}
\newcolumntype{R}[1]{>{\PreserveBackslash\raggedleft}p{#1}}
\newcolumntype{L}[1]{>{\PreserveBackslash\raggedright}p{#1}}
\usepackage{cite, url}
\usepackage{multicol, multirow}
\newcolumntype{Y}{>{\centering\arraybackslash}X}
\newcolumntype{k}{>{\arraybackslash}l}
\usepackage{tabularx}
\usepackage{nicefrac}

\title{Graph Attention Networks for Speaker Verification}

\name{Jee-weon Jung$^{1,2}$*, Hee-Soo Heo$^{2}$*, Ha-Jin Yu$^1$, and Joon Son Chung$^{2}$ \thanks{\hspace{-10pt}* These authors contributed equally to this work.}}
\address{$^1$School of Computer Science, University of Seoul, $^2$Naver Corporation}
\begin{document}
\ninept
\maketitle

\begin{abstract}
This work presents a novel back-end framework for speaker verification using graph attention networks. 
Segment-wise speaker embeddings extracted from multiple crops within an utterance are interpreted as node representations of a graph. 
The proposed framework inputs segment-wise speaker embeddings from an enrollment and a test utterance and directly outputs a similarity score. 
We first construct a graph using segment-wise speaker embeddings and then input these to graph attention networks. 
After a few graph attention layers with residual connections, each node is projected into a one-dimensional space using affine transform, followed by a readout operation resulting in a scalar similarity score. 
To enable successful adaptation for speaker verification, we propose techniques such as separating trainable weights for attention map calculations between segment-wise speaker embeddings from different utterances. 
The effectiveness of the proposed framework is validated using three different speaker embedding extractors trained with different architectures and objective functions. 
Experimental results demonstrate consistent improvement over various baseline back-end classifiers, with an average equal error rate improvement of 20\% over the cosine similarity back-end without test time augmentation. 
\end{abstract}
\begin{keywords}
Graph neural network, graph attention network, speaker verification
\end{keywords}

\section{Introduction}
\label{sec:intro}
Speaker recognition has received an increasing amount of attention as a means of authentication due to its non-invasive nature and easy accessibility.
Speaker recognition can be divided into closed-set or open-set settings, where most researchers focus on the latter problem. 
Open-set speaker recognition can be interpreted as a metric learning problem, in which voices must be mapped into a discriminative embedding space and the similarity scores between the embeddings are used to verify speakers.

There is a large body of recent literature on speaker verification that makes use of deep neural networks (DNNs) ~\cite{Nagrani17,snyder2017deep,snyder2018x,Jung2018AResult,ravanelli2018speaker,okabe2018attentive,snyder2019speaker}. 
The majority of the work learn similarity scores indirectly. 
For example, speaker embeddings ({\em i.e.} representations) are often trained via the classification task.
While the classification loss can learn separable embeddings, they are not designed to optimise embedding similarity. 
Recently, several works categorised as metric learning that directly optimise similarity metrics have proven success in speaker verification~\cite{Heo2019End-to-endVerification,zhang2018text,wan2018generalized,chen2020simple, He2020MomentumLearning,chung2020defence}. 
Triplet and generalized end-to-end (E2E)~\cite{zhang2018text,wan2018generalized} learn to minimize the cosine distance between intra-class utterances and maximize between different classes. 
More recent works~\cite{chen2020simple, He2020MomentumLearning, chung2020defence} have shown promising performance by training more discriminative embeddings via multiple negatives for every positive sample in training. 

Although the recent works in metric learning have shown success in speaker verification, separate back-end classifiers are often used to generate strong scoring functions. 
For instance, probabilistic linear discriminant analysis (PLDA) is a widely used back-end system that suits all front-end speaker embeddings represented in a fixed low-dimensional space ({\em e.g.} i-vector \cite{Dehak2010DiscriminativeVerification,Dehak2011Front-endVerification} and x-vector \cite{snyder2018x}). 
However, PLDA requires preprocessing techniques such as centering, linear discriminant analysis (LDA), and length normalization to achieve good performance. 
Due to such difficulty in the calibration of PLDA systems, DNN-based back-end systems have been proposed. 
For example, \cite{Jung2018AResult} proposes an end-to-end framework incorporating both embedding extraction and classification, and b-vector~\cite{Lee2014SpeakerFeatures} and relation module~\cite{sung2018learning} have been successfully applied to speaker recognition, outperforming the conventional back-end performance.
These techniques handle speaker verification as a binary classification or learn a specific metric for speaker verification using the networks. 
Despite extensive research in speaker verification back-ends, systems in the recent literature adopt a simple cosine similarity as a similarity score, since many studies report no improvement in performance when adopting an external back-end in place of cosine similarity~\cite{Jung2019RawNet:Verification,Wang2020SpeakerNetworks}. 

Two factors have contributed to the demise of the use of the back-end classification. 
Firstly, improvements in deep speaker embedding extractors have resulted in easily separable speaker embeddings using cosine similarity. 
Secondly, the use of test time augmentation (TTA)~\cite{Chung18a} has mitigated the difficulty of dealing with utterances of various duration in the test phase. 
With TTA, multiple segment-wise speaker embeddings (SSEs) with overlaps are extracted from an identical duration of utterance used in the training phase. 
The similarity score is derived by averaging pairwise similarity scores between SSEs. 

However, in the TTA framework, equal weights are given to all pairs of segments when calculating the final similarity score, although a specific segment pair may be more informative. 
In some cases ({\em e.g.} a noisy environment), we argue that it can be helpful to aggregate and dynamically model the relationship between the multiple segments, but this cannot be accomplished in the current framework based on the na\"ive averaging of the pairwise similarity scores. 
Therefore, we propose a novel graph neural network (GNN)-based back-end classification framework for speaker verification to overcome these limitations.

GNNs have been rapidly developed with powerful variants such as graph convolutional network (GCN)~\cite{Kipf2017Semi-SupervisedNetworks} and graph attention network (GAT)~\cite{PetarVelickovic2018GRAPHNETWORKS} in recent years. 
In speaker recognition, however, the use of graph neural networks have not been studied in the context of deep learning.
By modelling SSEs as nodes of a graph, we propose to use the GAT framework for {\em learning} a deep utterance-level similarity metric. 

Using the proposed GAT-based back-end framework, we argue that the underlying non-Euclidean data manifold between the SSEs can be leveraged, where this information is typically neglected by averaging in conventional frameworks. 
Sequence-agnostic aggregation methods that construct an utterance-level representation from multiple frame-level representations ({\em e.g.} attentive pooling \cite{Zhu2018Self-AttentiveVerification} and statistical pooling~\cite{snyder2018x}) show state-of-the-art performance.
By similar rationale, we argue that a graph structure instead of sequence models ({\em e.g.} recurrent neural networks) can be beneficial for improvement in a speaker verification system. 
Through a wide range of experiments on the VoxCeleb dataset~\cite{Nagrani17,Chung18a}, we demonstrate the superiority of the proposed method compared to various existing back-end systems. 
To the best of our knowledge, this study is the first to adapt a GNN for modelling speaker embeddings. 

\begin{figure*}[ht!]
    \centering
    \includegraphics[width=\textwidth]{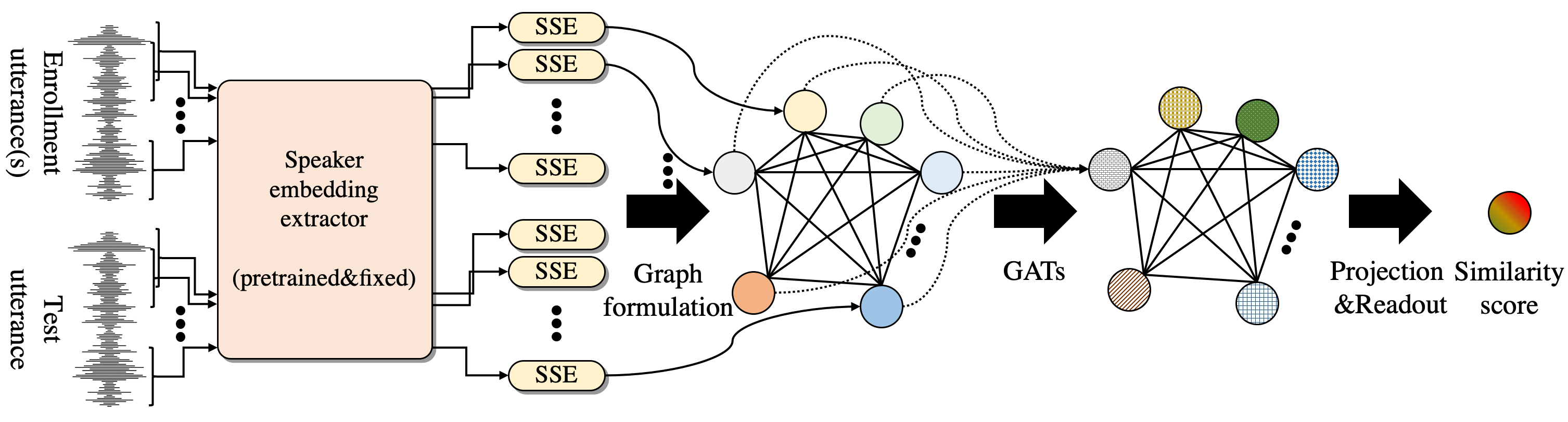}
    \caption{GAT-based back-end framework for speaker verification. Enrollment utterance(s) and a test utterance is input and a similarity score is calculated. (SSE: segment-wise speaker embedding, GAT: graph attention network)}
    \label{fig:gat_frame}
\end{figure*}

\begin{table}[t!]
  \centering
  \begin{tabularx}{\linewidth}{kYYY}
  \Xhline{1pt}
  

   Layer & Kernel size  & Stride& Output shape   \\
  \Xhline{1pt}
  Conv1 & $3 \times 3 \times 32$ & $1\times1$& $L \times 64\times 32$ \\
  \hline
  Res1 & $3\times 3 \times 32$ & $1\times1$ & $ L \times64 \times 32$ \\
  \hline
  Res2 & $3\times 3\times 64$ &$2\times2$ & $ \nicefrac{L}{2} \times32 \times 64$ \\
  \hline
  Res3 & $3\times 3 \times 128$ &$2\times2$ & $ \nicefrac{L}{4} \times16 \times 128$\\
  \hline
  Res4 & $3\times 3\times 256$ &$2\times2$ & $ \nicefrac{L}{8} \times8 \times 256$\\
  \hline
  Flatten & - & - & $ \nicefrac{L}{8} \times 2048$\\
  \hline
  ASP & - & - & $ 4096$ \\
  \hline
  Linear & $ 512 $ & - & $ 512 $ \\
  \Xhline{1pt}
  \end{tabularx}
  \label{table:model}
  \caption{Model architecture for the {\bf ResNet-S} model. {$\mathbf{L}$}: length of input sequence, {\bf ASP}: attentive statistics pooling.}
\end{table}

\section{Speaker embedding extractor}
\label{sec:spk_embd}
With advances in deep learning and the availability of large-scale datasets, the use of DNNs for speaker verification has become increasingly popular.
In particular, variants of the residual networks~\cite{He2016IdentityNetworks} are widely used for speaker verification. 
We use three pretrained deep speaker embedding extractors, also variants of residual networks, to evaluate various back-end classifiers. 
All three systems extract frame-level representations using convolutional blocks with residual connections, which are then aggregated into a single segment-level representation. 
To verify the effectiveness of our method across a range of embedding extractors, we experiment with three embedding extractors that have different architecture and are trained with different loss functions. 

The first variant is identical to that of the authors' previous work~\cite{chung2020defence}. 
With only 1.4 million weight parameters, it is one of the lightest and the fastest networks introduced in the recent literature in speaker recognition. 
This model is trained using angular prototypical loss, which is an angular variant of the prototypical networks~\cite{snell2017prototypical}. 
We refer to this model as {\bf ResNet-F}.
The second model is the original ResNet-34, which includes 22.0 million parameters, trained using a hard negative loss. 
Detailed model architecture and the loss function is described in~\cite{Heo2019End-to-endVerification}. 
We refer to this model as {\bf ResNet-O}.
The final model is a modified ResNet-34 suitable for speaker verification which has 8.0 million parameters \cite{heo2020clova}. 
In particular, the channels of each residual block are reduced by half, compared to the original ResNet-34. 
However, the strides at every layer up to the second residual block have been removed, which helps the model to retain the input resolution in the first few hidden layers at the cost of increased computations.
Attentive statistics pooling (ASP)~ \cite{okabe2018attentive} is used to aggregate temporal frames within SSEs.
In the ASP layer, the channel-wise weighted standard deviation is calculated in addition to the weighted mean.
This model is trained using the angular prototypical loss and the softmax cross-entropy loss simultaneously. Table \ref{table:model} describes the detailed architecture of final model. This model has the slowest inference time of the three, and we refer to this model as {\bf ResNet-S}. The pre-trained weight parameters of ResNet-F and ResNet-S architectures are publicly available.\footnote{ \url{https://github.com/clovaai/voxceleb\_trainer}.} 

\section{Back-end classifiers}
\label{sec:back-end}
This section describes the baseline back-end systems and introduces the proposed GNN-based back-end classifier.
\subsection{Baselines}
\label{ssec:backend_baseline}
Once speaker embeddings are extracted from the enrollment and the test utterance, the back-end system measures the similarity score between them. 
Support vector machines \cite{Cumani2011FastSpace} and PLDA are widely adopted as back-end systems, used in conjunction with i-vectors \cite{Dehak2011Front-endVerification}. 
In particular, many variants expanding the PLDA such as discriminative PLDA \cite{Burget2011DiscriminativelyVerification} and neural PLDA \cite{Ramoji2020NPLDA:Verification} have been explored. 

DNN-based back-ends systems are an active field of study. 
Lee \textit{et al.}~\cite{Lee2014SpeakerFeatures} proposed to use a concatenation of element-wise addition, subtraction, and multiplication as a new representation (i.e., b-vector) and adopt a few fully-connected layers to perform back-end speaker verification, which has been extended in \cite{Heo2016AdvancedVerification, Jung2019RawNet:Verification}. 
However, the effectiveness of aforementioned systems have been reported to decrease with the development of more discriminative DNN-based speaker embeddings \cite{Jung2019RawNet:Verification, Wang2020SpeakerNetworks}. 
Moreover, conventional back-end systems cannot fully utilize multiple SSEs that are used in recent works, because these systems are designed for a scheme where one speaker embedding is extracted per utterance. 
Hence, the development of novel back-end classifiers that can model neglected data manifold between multiple SSEs is required to fully exploit the potential.

\subsection{GAT-based back-end classifier}
\label{ssec:GAT}
In this section, we introduce the proposed GAT-based speaker verification framework. 
To construct a graph using SSEs, we interpret each SSE as a node of a graph.
\subsubsection{Model definition}
\label{sssec:gat_arch}
Since each node is an SSE, we do not predetermine any connections between them, but rather adopt an attention mechanism to emphasize and de-emphasize such connections. 
Due to this property, we argue that the proposed GAT-based framework overcomes the limitations of the existing TTA framework that typically average SSEs to form a single vector representation, ignoring the data manifold between them. 

Using the pretrained embedding extractors introduced in Section~\ref{sec:spk_embd}, we first extract SSEs from both the enrollment and the test utterances. 
In the scenarios where multiple utterances are used for speaker enrollment, one can use the element-wise average of different utterances' SSEs. 
Next, a graph is formed using SSEs of both utterances. 
For a configuration that extracts ten SSEs per utterance, for example, the formed graph will have twenty nodes. 
Note that the graph we use has all possible edges including self-connections, and the weight of each edge is determined using an attention mechanism. 
Formally, let $x$ be an SSE, $x \in \mathbb{R}^d$ where $d$ refers to the dimensionality of an SSE.  
We denote the formed input graph as $\mathcal{G}$, $\mathcal{G} \in \mathbb{R}^{n \times d}$ where $n$ is the number of SSEs extracted from two utterances (enrollment and test). 

Once a graph is formed, we use several graph attention layers with residual connections \cite{Pham2017ColumnClassification}. 
We then project the output of the last graph attention layer $\mathcal{G}'$, $\mathcal{G}' \in \RR^{n \times d'}$ where $d'$ is the output dimensionality, into a scalar using an affine transform ({\em i.e.} dense layer). 
Finally, we perform readout by averaging the nodes to derive the similarity score of the corresponding pair of utterances. 
In the training phase, we calculate contrastive \cite{Chen2020ARepresentations, He2020MomentumLearning} and hard negative loss \cite{Heo2019End-to-endVerification} within each mini-batch to update the weight parameters. 
In the test phase, we use the similarity score directly to perform speaker verification. 
Figure~\ref{fig:gat_frame} illustrates the overall framework of the proposed GAT-based back-end classifier. 

\subsubsection{GAT-based framework}
\label{sssec:gat}
In this section, we describe the detailed operations inside the proposed GAT-based framework. 
The goal of the proposed system is to calculate a similarity score from a graph $\mathcal{G}$ using a learnable function $GAT(\mathcal{G}):\mathbb{R}^{n\times d} \rightarrow \mathbb{R}$ \cite{Kipf2017Semi-SupervisedNetworks}.
Let $\boldsymbol{h}^{(k)}_{u}$ be the representation of node $u$ after $k$ graph attention layers and $\boldsymbol{h}^{(k)}$ be a set of nodes. 
$\boldsymbol{h}^{(0)}_{u}$ is the representation of node $u$ from the input graph.
Thus, $\mathcal{G} =\boldsymbol{h}^{(0)}$ and $\mathcal{G}' =\boldsymbol{h}^{(K)}$ where $K$ is the number of graph attention layers. 
The proposed GAT-based framework introduced in Section~\ref{sssec:gat_arch} is denoted as:

\begin{equation}
\begin{split}
    \label{GAT_fn}
    GAT(\mathcal{G})&=\frac{1}{n}\sum_{u \in \mathcal{G}'}\boldsymbol{h}^{(K)}_{u}W_{out},\\
    \boldsymbol{h}^{(k)}&=f(\boldsymbol{h}^{(k-1)}),
\end{split}
\end{equation}

\noindent where $n$ is the number of nodes, $W_{out} \in \mathbb{R}^{d' \times 1}$ is the trainable parameters for the affine transform, and $f$ is a graph attention layer. 
Note that in \eqref{GAT_fn}, addition of bias is omitted for brevity. 

The propagation of each node $\boldsymbol{h}^{(k)}_{u}$ is defined as a combination of the previous hidden representation and aggregated information. 

\begin{equation}
   \label{GAT_hk}
   \boldsymbol{h}^{(k)}_{u} = \sigma(MLP_{\phi}(\boldsymbol{m}^{(k)}_{u}) + MLP_{\psi}(\boldsymbol{h}^{(k-1)}_{u})),
\end{equation}

\noindent where $\sigma(\cdot)$ is non-linear activation function, $\boldsymbol{m}^{(k)}_{u}$ is the aggregated information for node $u$, and $MLP_{\phi}: \mathbb{R}^{d^{k-1}} \rightarrow \mathbb{R}^{d^{k}}$ and $MLP_{\psi}: \mathbb{R}^{d^{k-1}} \rightarrow \mathbb{R}^{d^{k}}$ are multi-layer perceptrons (MLPs) with a set of learnable parameters $\phi$ and $\psi$, respectively. 
$MLP_{\psi}$ denotes the residual connection in GNNs where the transformation is applied to deal with change in dimensionality between graph attention layers, similar to the downsampling residual connection in residual networks.  
With \eqref{GAT_hk}, the representation of each node is iteratively applied.
For aggregating the information from neighborhoods, we apply an attention strategy~\cite{Bahdanau2015NeuralTranslate, PetarVelickovic2018GRAPHNETWORKS}.

\begin{equation}
    \label{GAT_na}
    \boldsymbol{m}^{(k)}_{u}=\sum_{v \in \mathcal{N}(u) \cup \{ u \} }\alpha^{(k)}_{v,u}\boldsymbol{h}^{(k-1)}_{v}, 
\end{equation}

\noindent where $\mathcal{N}(u)$ is the neighborhood set for node $u$, and $\alpha^{(k)}_{v,u}$ is the attention weight from the node $v$ to the node $u$. 
We configure the neighborhood set for node $u$ as all nodes on the graph, including itself.
The attention weight is calculated in a symmetrical form using a separate MLP and softmax function in which we argue that it enables individual nodes to aggregate meaningful nodes.

\begin{equation}
    \label{GAT_sm}
    \alpha^{(k)}_{u,v}=\frac{exp(g^{(k)}(\boldsymbol{h}^{(k-1)}_{u}, \boldsymbol{h}^{(k-1)}_{v}))}{\sum_{w \in \mathcal{N}(u) \cup \{ u \}}  exp(g^{(k)}(\boldsymbol{h}^{(k-1)}_{u}, \boldsymbol{h}^{(k-1)}_{w}))  }, 
\end{equation}
\noindent where $exp(\cdot)$ is the exponential function, $w$ is also a node index, and MLP function $g^{(k)}(\cdot, \cdot)$ is defined depending on the type of nodes as bellow:

\[
    g^{(k)}(\boldsymbol{h}^{(k-1)}_{u}, \boldsymbol{h}^{(k-1)}_{v})= \\ 
\begin{cases}
    MLP_{\theta1 }(\boldsymbol{h}^{(k-1)}_{u} \otimes \boldsymbol{h}^{(k-1)}_{v}),& \text{if } u \in \mathcal{U}(v)\\
    MLP_{\theta2 }(\boldsymbol{h}^{(k-1)}_{u} \otimes \boldsymbol{h}^{(k-1)}_{v}),              & \text{otherwise}
\end{cases}
\]

\noindent where $\otimes$ is element-wise multiplication making symmetric attention weights, and $MLP_{\theta 1,2}: \mathbb{R}^{d^{k-1}} \rightarrow \mathbb{R}$. 
In addition, $\mathcal{U}(v)$ is a set of nodes (or SSEs) of the utterance from which the node $v$ is extracted.
In other words, if $u \in \mathcal{U}(v)$, node $u$ and $v$ are from the same utterance. 
The choice of element-wise multiplication is inspired by the b-vector technique, proposed for speaker recognition~\cite{Lee2014SpeakerFeatures}. 
We expect that this technique, which separately calculates attention weights for SSEs from different utterances, enables effective training for comparing SSEs in different utterances or aggregating SSEs within the same utterance. 

\subsubsection{Loss functions}
\label{sssec:loss}
We use two loss functions to train the proposed back-end GAT framework: contrastive and hard negative. 
The contrastive \cite{Chen2020ARepresentations, He2020MomentumLearning} loss calculates the distance of a pair of samples, which has been widely used for recent self-supervised learning frameworks. 
Each mini-batch comprises two utterances for $M$ speakers, thus, resulting in $2M$ utterances in total. 
We denote a set of SSEs of each mini-batch as $\boldsymbol{x}_{i,j}|1\leq i\leq M, j \in \{ 1,2\}$. 
Note that one SSE set $\boldsymbol{x}_{i,j}$ contains multiple ({\em e.g.} ten) SSEs extracted from an utterance. 
Using the only pair composed of the identical speaker as target and all the rest as non-target, a contrastive loss is calculated by applying a softmax non-linearity where the target is the numerator and the summation of non-targets are the denominator. 
Then, a categorical cross-entropy loss is calculated. 
Formally, the contrastive loss that we use for training the GAT network is defined as:
\begin{equation}
    \label{lc}
    \vspace{3pt}
    \mathcal{L}_{c}
    =-\frac{1}{M}\sum^{M}_{i=1}log\frac{exp(GAT(\mathcal{G}_{\boldsymbol{x}_{i,1},  \boldsymbol{x}_{i,2}}))}
    {\sum^{M}_{k=1}
    exp(GAT(\mathcal{G}_{\boldsymbol{x}_{i,1}, \boldsymbol{x}_{k,2}}))}, 
\end{equation}
where $\mathcal{G}_{\boldsymbol{x}_{i,1}, \boldsymbol{x}_{i,2}}$ is the graph formulated using two utterances $\boldsymbol{x}_{i,1}$ and $\boldsymbol{x}_{i,2}$ that belongs to an identical speaker. 
The contrastive loss minimizes intra-class variance and maximizes inter-class variance simultaneously. 
Because it adopts softmax non-linearity, it also contains the idea of hard negative mining. 

The hard negative \cite{Heo2019End-to-endVerification} loss function has been proposed to reinforce the concept of hard negative mining for a categorical cross-entropy loss in the speaker recognition field. 
We adopt the concept of hard negative loss to the contrastive loss defined based on the categorical cross-entropy loss in \eqref{lc}.

\begin{equation}
    \label{lh}
    \vspace{3pt}
    \mathcal{L}_{h}
    =-\frac{1}{M}\sum^{M}_{i=1}log\frac{exp(GAT(\mathcal{G}_{\boldsymbol{x}_{i,1},  \boldsymbol{x}_{i,2}}))}
    {\sum_{\boldsymbol{h} \in \mathcal{H}_i}
    exp(GAT(\mathcal{G}_{\boldsymbol{x}_{i,1}, \boldsymbol{h}}))}, 
\end{equation}

\noindent where $\mathcal{H}_i$ is the set of the top $H$ utterances with large $GAT(\mathcal{G}_{\boldsymbol{x}_{i,1},  \boldsymbol{x}_{k,2}})\\|_{k \neq i,1 \leq j \leq M}$.
The main difference from \eqref{lc} to \eqref{lh} is that the denominator part of \eqref{lc} has been replaced with the hard negative set. 

\section{Experiments}
\label{sec:exp}
\subsection{Dataset}
\label{ssec:db}
We train the models using the development set of VoxCeleb2 \cite{Chung18a}. 
It comprises over a million utterances from 5,994 speakers with an average duration of 7.2 seconds. 
We report performances of all systems in terms of equal error rates (EERs) using the original VoxCeleb1 \cite{Nagrani17} test set which includes 37,720 trials from 40 speakers that do not appear in the training set. 
All utterances are provided in a single-channel PCM with a sampling rate of 16 kHz. 

\subsection{Implementation details}
\label{ssec:exp_config}
We conduct all experiments using the PyTorch framework, on the NAVER Smart Machine Learning platform~\cite{Sung2017NSML:Models}. 
We evaluate the performances of cosine similarity, PLDA, b-vector, and relation modules as the back-end baseline. 
The performances of all back-end classifiers are evaluated using three different speaker embedding extractors.

\newpara{Baselines.}
To adopt the PLDA back-end, we apply LDA and length normalization. 
With the LDA application, the performances are evaluated by reducing the 512-dimensional embedding vector to 256 and 128 dimensions. 
To extract the b-vector from the two embedding vectors, the concatenation of two vectors utilized in \cite{Jung2019RawNet:Verification} is applied in addition to element-wise operations such as addition, multiplication, and subtraction. 
The b-vectors are classified into two classes using an MLP with three hidden layers, where one class indicates that two utterances are from the same identity and another means the different speakers~\cite{Lee2014SpeakerFeatures}. 
The relation module is also implemented as an MLP including three hidden layers and trained using contrastive loss.

\newpara{Proposed method.}
The GAT framework described in section \ref{sssec:gat} comprises three layers, and all MLPs ($MLP_{\phi}, MLP_{\psi}$, and $MLP_{\theta}$) included in each graph attention layer contain one linear operation.
All networks including that of the conventional back-ends are trained for 200 epochs using the Adam optimizer.
Note that the embedding extractor is frozen while the back-end network is trained. 
We use an initial learning rate of 0.001 and annealed it using a cosine function without a restart \cite{Loshchilov2017SGDR:Restarts}.
We use a mini-batch size of 350.
Without batch normalization, only drop-out is applied at a rate of 0.2 to the input.
Weight decay with $10^{-4}$ ratio is applied to all networks.

\begin{table}[t!]
  \centering
  \begin{tabularx}{\linewidth}{lYYY}
  \Xhline{1pt}
   Back-end type \& & \multicolumn{3}{c}{Speaker embedding extractor}\\
  \cmidrule{2-4}
   hyper-parameters & ResNet-F & ResNet-O & ResNet-S\\
  \Xhline{1pt}
  CosSim w/o TTA & 2.57  & 1.89 & 1.38    \\ 
  CosSim w/ TTA & 2.17     & 1.73    &  1.17    \\ 
  \Xhline{0.5pt}
   PLDA w/o LDA &     4.24     & 2.83    & 1.69         \\ 
   PLDA w/ LDA (256) &  3.99    & 2.67    & 1.88         \\  
   PLDA w/ LDA (128) &  3.72    & 2.88   & 2.18         \\ 
   \Xhline{0.5pt}
   b-vector ($\otimes, \oplus$) & 2.96 & 1.85    &   1.51      \\ 
   b-vector ($\otimes, \oplus, \ominus$) & 2.22         & 1.65    &     1.36    \\ 
   b-vector ($\otimes, \doubleplus$)&  2.86        & 1.81    &  1.44       \\
   \Xhline{0.5pt}
   Relation module & 2.24     & 2.20    & 2.14    \\ 
   \Xhline{0.5pt}
   GAT ($\mathcal{L}_{c}$) & 2.09     & 1.62    & 1.17    \\ 
   GAT ($\mathcal{L}_{h}$) & \textbf{2.06}     & \textbf{1.52}    & \textbf{1.09}    \\
  \Xhline{1pt}
  \end{tabularx}
  \caption{Comparison of the proposed GAT-based framework using two loss functions with various conventional back-end classifiers. All results are reported using three different speaker embeddings to increase the reliability. The values in boldface describe the best classifier performance for each speaker embedding. Performance reported in EERs (\%). Cosine similarity without TTA is used as the baseline. {\bf TTA}: test time augmentation. $\otimes$, $\oplus$, $\ominus$, and $\doubleplus$: element-wise multiplication, addition, subtraction, and vector concatenation, respectively, for b-vector construction. $\mathcal{L}_{c}$: contrastive loss and $\mathcal{L}_{h}$: hard negative mining loss. }
  \label{tab:gat_embd}
\end{table}
\vspace{-3pt}
\subsection{Results}
\label{ssec:res}
Table~\ref{tab:gat_embd} reports the result of the experiments using various back-end classifiers including the proposed GAT-based back-end framework. 
The performance of each back-end classifier is reported using three different embedding extractors: ResNet-F, ResNet-O, and ResNet-S. 
In case of the GAT-based framework, we report performances using two different loss functions. 
A cosine similarity classifier without TTA is used as the baseline for each embedding extractor. 

Our results on TTA and PLDA coincide with those reported in the literature. 
TTA outperforms the corresponding baseline for all three types of speaker embeddings. 
PLDA, regardless of the application of LDA and the configuration of reduced dimensionality, degrades the performance for all speaker embeddings.  

Results on the b-vector and relation module back-ends also meet previously reported tendency, in which the extent of performance increase decreases as more discriminative front-end speaker embedding is used. 
b-vector back-end using all three element-wise operations consistently outperforms the baseline, however, a relative reduction of EER is the biggest in ResNet-F (2.57\%$\rightarrow$2.22\%) and smallest in ResNet-S (1.38\%$\rightarrow$1.36\%).
For the other two b-vector configurations, only ResNet-O outperforms cosine similarity with a very small margin. 
Relation module back-end also shows the same tendency where performance improves only in case of ResNet-F. 
We interpret these results as that the TTA procedure and the advanced embedding extractors have replaced most of the functions of the conventional back-end classifiers.

The proposed GAT-based framework outperforms consistently outperforms corresponding baselines, regardless of the loss function configuration. 
Even when compared with cosine similarity with TTA, which outperforms other conventional back-end classifiers in most cases, the proposed GAT-based frameworks outperforms in all configurations except the case where SSEs from ResNet-S is used with GAT trained with contrastive loss (GAT ($\mathcal{L}_{c}$)). 
Among two loss functions, using hard negative loss GAT ($\mathcal{L}_{h}$) demonstrates the best performance for all three embedding extractors. 
The average relative error reduction rate is 20\% where ResNet-S demonstrates the largest improvement with 21\% compared with cosine similarity without TTA. 

\section{Conclusion}
\label{sec:conclusion}
We introduced a GAT-based framework which ingests SSEs extracted from a pair of utterances and outputs a similarity score for speaker verification. 
By interpreting each SSE as a node of a graph and constructing a graph using SSEs from the enrollment and the test utterance, we were able to successfully adapt GATs with residual connections which is then projected into a one-dimensional space representing similarity score between two utterances. 
Despite rapid advances in DNN-based front-end embedding extractors, the proposed framework demonstrates consistent improvement over the cosine similarity baseline, whereas conventional classifiers such as PLDA and b-vector tend to degrade the performance. 
The best performing system demonstrated an EER of 1.09\% on the original VoxCeleb evaluation protocol. 
Based on the successful adaptation of GATs for speaker verification, as our future work, we plan to expand this work into a graph classification scheme. 

\clearpage
\ninept
\bibliographystyle{IEEEbib_mod}
\bibliography{shortstrings,mybib,vgg_local,vgg_other,references_jung_mendeley_local}
\end{document}